\documentclass[journal=jpclcd,manuscript=letter,layout=traditional]{achemso}
\usepackage{chemformula}
\usepackage[T1]{fontenc}

\usepackage{epsfig,bm}

\author{Francesco Cordero}
\email{francesco.cordero@ism.cnr.it}
\author{Floriana Craciun}
\email{floriana.craciun@ism.cnr.it}
\affiliation[ARTOV]
{Istituto di Struttura della Materia-CNR (ISM-CNR), Area della Ricerca di Roma - Tor Vergata, Via del Fosso del Cavaliere 100, I-00133 Roma, Italy}
\author{Francesco Trequattrini}
\affiliation[Fis]
{Dipartimento di Fisica, Universit\`{a} di Roma "La Sapienza", p.le A. Moro 2, I-00185 Roma, Italy}
\alsoaffiliation[ARTOV]
{Istituto di Struttura della Materia-CNR (ISM-CNR), Area della Ricerca di Roma - Tor Vergata, Via del Fosso del Cavaliere 100, I-00133 Roma, Italy}
\author{Patrizia Imperatori}
\author{Anna Maria Paoletti}
\author{Giovanna Pennesi}
\email{giovanna.pennesi@ism.cnr.it}
\affiliation[MLIB]
{Istituto di Struttura della Materia-CNR (ISM-CNR), Area della Ricerca di Roma 1, Via Salaria, Km 29.300, I-00015 Monterotondo Scalo, Roma, Italy}

\title[Polar and Antiferrodistortive Modes in MAPbI$_3$]
{Competition between Polar and Antiferrodistortive Modes and Correlated Dynamics of the Methylammonium Molecules in 
MAPbI$_3$ from Anelastic and Dielectric Measurements}

\begin{document}

\begin{tocentry}
\includegraphics[width=5 cm]{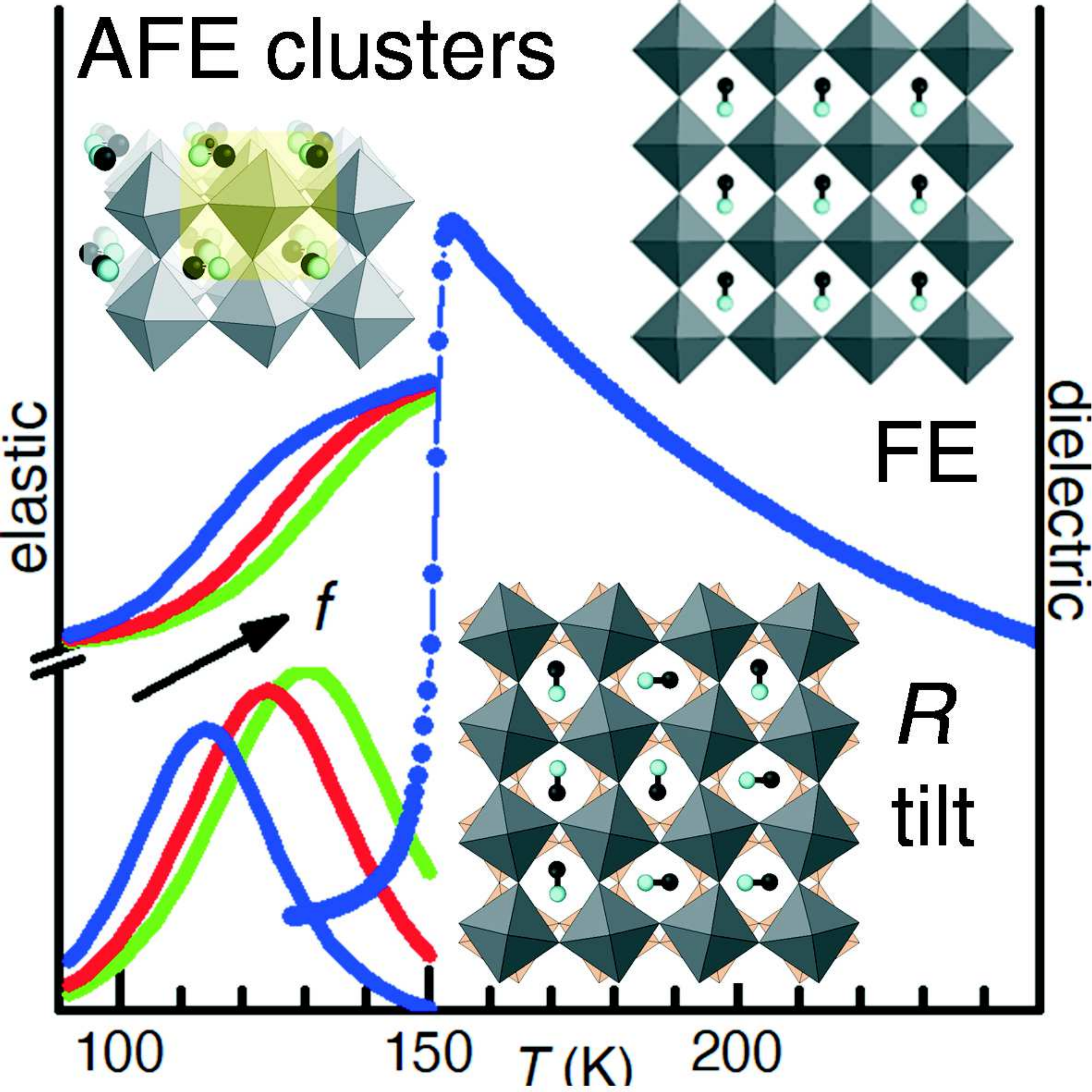}
\end{tocentry}

\begin{abstract}

The mechanisms behind the exceptional photovoltaic properties of the
metallorganic perovskites are still debated, and include a ferroelectric
(FE) state from the ordering of the electric dipoles of the organic molecules.
We present the first anelastic
(complex Young's modulus) and new dielectric measurements on
CH$_{3}$NH$_{3}$PbI$_3$, which provide new insight on the reorientation
dynamics of the organic molecules, and the reason why they do not form
a FE state. The permittivity is fitted within the
tetragonal phase with an expression that includes the coupling between FE
and octahedral tilt modes, indicating that the coupling is competitive
and prevents FE ordering. The onset of the orthorhombic phase is accompanied
by sharp stiffening, analogous to the drop of permittivity, due to the
hindered molecular dynamics. On further cooling, an intense
anelastic relaxation process without a dielectric counterpart
suggests the reorientation of clusters of molecules with strong
antiferroelectric correlations.

\end{abstract}


The metallorganic lead-halide perovskites with formula MAPbX$_{3}$ (MA =
methylammonium ion CH$_{3}$NH$_{3}$, X\ = Cl, Br, I)\ have been studying
since at least the 1980s,\cite{WKM85,PW87,OMS92} but started attracting
enormous attention in the last years, thanks to their exceptional
photovoltaic properties,\cite{Par13} which allow inexpensive solar cells
with efficiencies of over 20\% to be constructed. Some important issues are
still debated, {\it e.g.} the manner of dealing with their scarce
stability, and the mechanisms responsible for their excellent photovoltaic
properties.\cite{HYS17} Closely connected with the latter is the often invoked
occurrence of ferroelectricity at room temperature. Indeed, an internal
field arising from a spontaneous polarization would be
highly beneficial to the transport of the photocarriers from the bulk to the
electrodes,\cite{FBB14} and also the walls separating ferroelectric
(FE) domains have
been proposed to be potentially beneficial for the generation and mobility
of the photocarriers.\cite{FBB14,LZK15} Various indications of
ferroelectricity or antiferroelectricity\cite{SKM16} (AFE) switchable to FE
have been reported, such as Piezo Force Microscopy images of
switchable FE domains\cite{KYZ14} and $P-E$ loops,\cite{SMK13c,SKM16,RBM17}
while in the related metallorganic perovskites based on MnCl$_{3}$ instead
of PbI$_{3}$ octahedra, even a sizeable piezoelectric response has been
measured.\cite{YLZ17} On the other hand, there are also several failed
attempts to reveal hard indications of a room temperature FE state, often
with alternative explanations for the successful observations,\cite%
{BEJ15,FXS15,HYS17,SBM17} so that the issue of ferroelectricity in the
organic halide perovskites is considered open.\cite{HYS17} Numerous are also
the predictions of FE states at room temperature from first principle
calculations,\cite{FBB14} even though it has been objected that the
orientational order of the polar MA molecules would be thermally disrupted
starting from very low temperature.\cite{FDS15}

We report anelastic (complex Young's modulus) and dielectric measurements
of MAPbI$_3$, which in combination
reveal new features on the reorientation dynamics of the MA molecules, and
the hindrance of their FE ordering by coupling with the tilt
modes of the PbI$_{6}$ octahedra.


MAPbI$_3$ was prepared by a one-step process at room temperature in an
environmentally friendly solvent with ultrasound assistance, as described
in the Supporting Information (SI). X-ray diffraction on the microcrystalline
powder showed a single tetragonal phase without contaminants like PbI$_2$ or hydrates
of the compound. The samples were pressed as discs and thin bars (SI).

The complex Young's
modulus, $E=E^{\prime }+iE^{\prime \prime }$, whose reciprocal is the compliance
$s=s^{\prime }-is^{\prime \prime }$, the mechanical analogue of the dielectric
susceptibility, was measured by exciting the
flexural modes of the bars suspended in vacuum on thin wires.
The resonance frequencies are $f \propto \sqrt{E^{\prime }}$,\cite{NB72}
and therefore, choosing a reference temperature $T_{0}$, one can plot
$E\left( T\right) /E\left( T_{0}\right) =$ $f^{2}\left( T\right) /f^{2}\left(
T_{0}\right) $. The
elastic energy loss $Q^{-1}=$ $E^{\prime \prime }/E^{\prime }=$ $s^{\prime
\prime }/s^{\prime }$ was measured from the free decay or from the resonance
curve under forced vibration.
The complex dielectric permittivity, $\epsilon =$ $\epsilon ^{\prime
}-i\epsilon ^{\prime \prime }$, with losses $\tan \delta =\epsilon ^{\prime
\prime }/\epsilon ^{\prime }$, was measured on the discs or fragments of the bars.
Experimental details and some information on anelasticity are found in the SI.

\begin{figure}[tbh]
\includegraphics[width=8.25 cm]{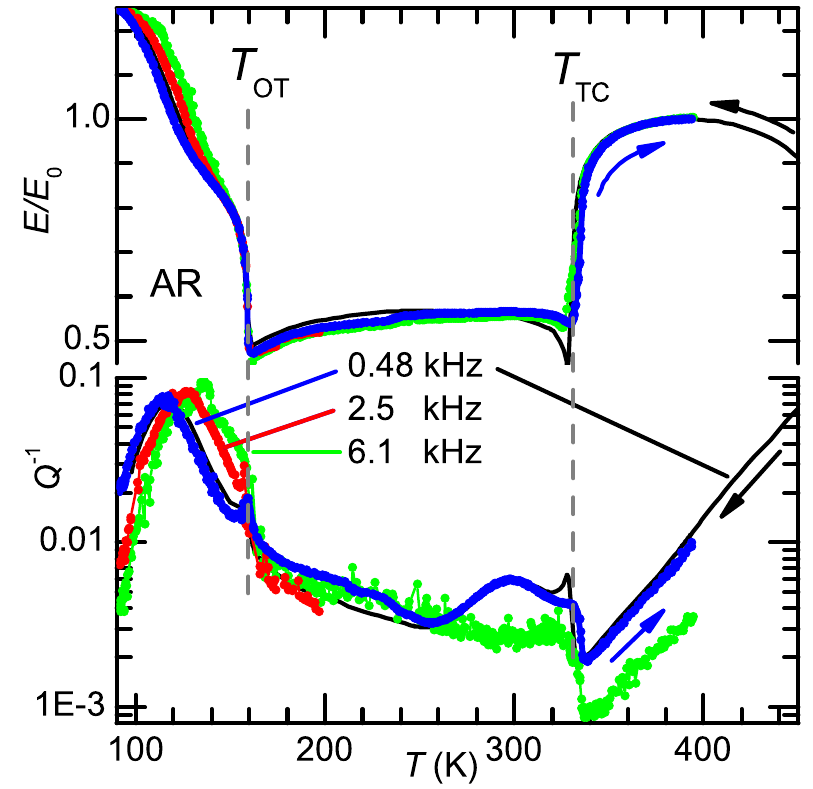}
\caption{Real part (upper panel) and losses (lower panel) of the Young's
modulus of MAPbI$_{3}$, measured at three frequencies during a same run.}
\label{fig-anel}
\end{figure}

\begin{figure}[h]
\includegraphics[width=12 cm]{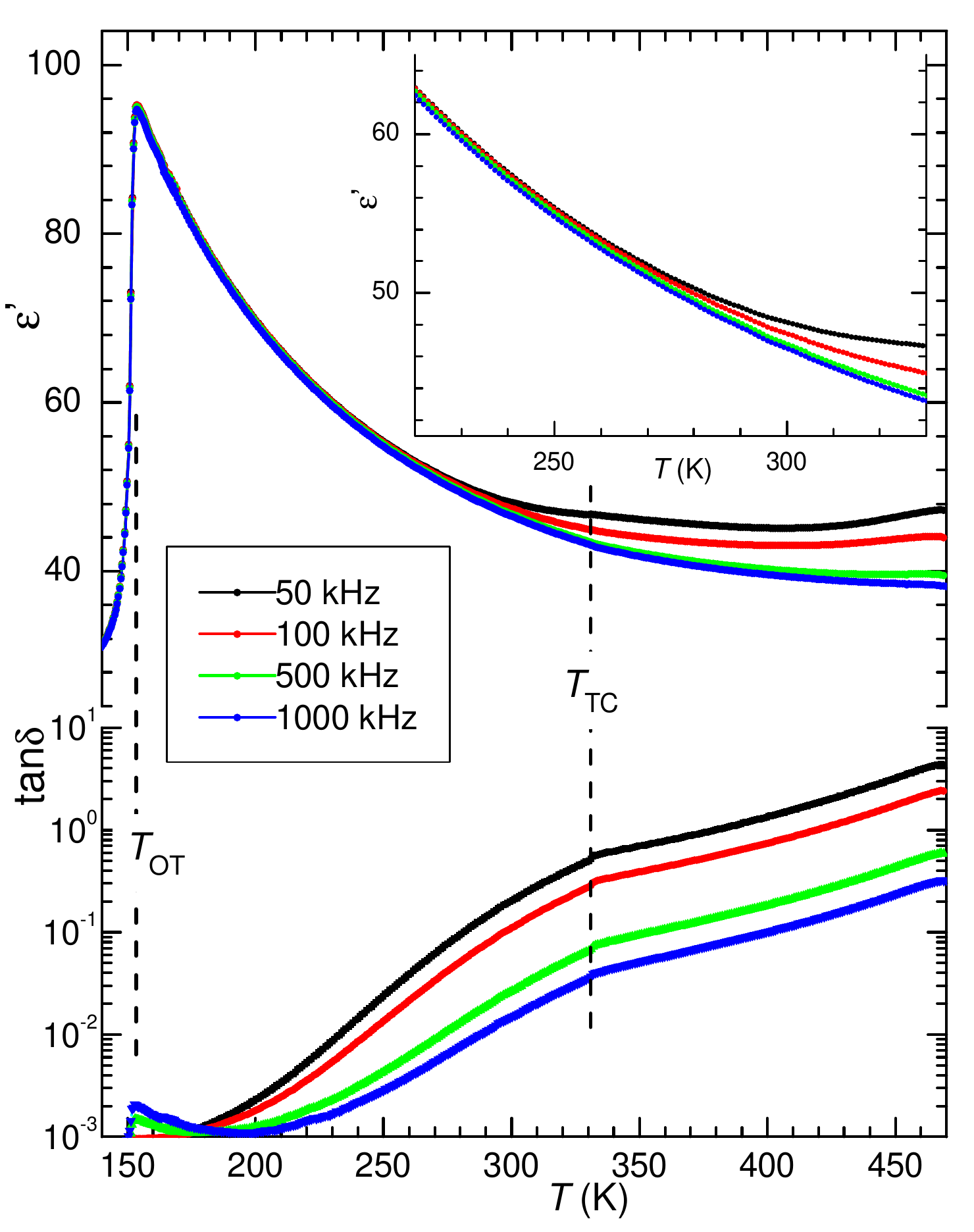}
\caption{Complete set of the $\protect%
\epsilon ^{\prime }$ and $\tan \protect\delta $ curves from which the curve
at 1~MHz has been used for fitting.}
\label{fig-S3}
\end{figure}

Figure \ref{fig-anel}
presents the anelastic spectrum of a MAPbI$_{3}$ bar measured during cooling
and subsequent heating, at $\sim 0.48$, 2.5 and 6.1~kHz.
The upper panel shows the real part of the complex Young's
modulus normalized to the maximum
value $E_{0}$ in the cubic (C) phase, and the lower panel the elastic energy
loss (for the structural phases see \emph{e.g.} Refs. \citenum{SMK13c,WHG16}).
There are two sharp steps
in $E^{\prime }$ at the transition temperature $T_{\mathrm{TC}}\simeq 328$~K
to the tetragonal (T) phase, with $2-5$~K of hysteresis between heating and
cooling, and at the transition to the orthorhombic (O) phase below $T_{%
\mathrm{OT}}\simeq 163$~K, with $<3$~K of hysteresis. A steplike softening,
like that at $T_{\mathrm{TC}}$, is expected from the standard Landau theory
both for tilt and FE phase transitions. In fact, the softening is due to
the coupling between strain $\varepsilon $ and the order parameter, tilt
angle $Q$ or polarization $P$, which in both cases are described by a term $%
\propto \varepsilon Q^{2}$ or $\varepsilon P^{2}$ in the free energy.\cite%
{Reh73} A typical example of perovskite showing both transitions is
PbZr$_{1-x}$Ti$_x$O$_3$ (PZT),
where the polar interactions producing ferroelectricity are much stronger
than the lattice mismatch between the anharmonic Pb$-$O framework and the
rigid (Ti/Zr)O$_{6}$ octahedra, which cause octahedral tilting.\cite{CCT14c}
Therefore, in PZT the Curie temperature $T_{\mathrm{C}}$ is larger than the
tilt transition temperature $T_{\mathrm{T}}$ and also the steplike softening
at $T_{\mathrm{C}}$ is larger than that at $T_{\mathrm{T}}$. In the
metallorganic halides the opposite is true: they exhibit clear multiple tilt
transitions, but the FE state at room temperature is, if any, weak,
unconventional and controversial, except for few exceptions.\cite{YLZ17,PLT17} The
comparison between the elastic and dielectric responses, the latter
practically unperturbed around $T_{\mathrm{TC}}$ (Fig. \ref{fig-S3} of SI and Refs. %
\citenum{OMS92,GKB17,ABG17b}), leaves no doubt that the transition at this
temperature involves only octahedral tilting.

The steplike stiffening, instead of softening, below $T_{\mathrm{OT}}$ is a
less conventional behavior. Few cases have been reported, where stiffening
occurs at a transition involving octahedral tilting; in BaCeO$_{3}$\
and SrZrO$_{3}$\ this occurs passing from a orthorhombic or tetragonal tilted high
temperature phase to the orthorhombic $Pnma$ phase,\cite{KHC09,ZKS10b} with
the same tilt pattern of MAPbI$_{3}$ ($a^{-}b^{+}a^{-}$ in Glazer's
notation, see SI), but the stiffening could not be explained within the usual
Landau treatment.\cite{KHC09,ZKS10b} Also PbZrO$_{3}$ stiffens when passing
from the untilted or disorderly tilted FE $R3m$ phase to the AFE $Pbam$
tilted ($a^{-}a^{-}c^{0}$) phase, but in that case the stiffening can be
explained by the loss of the piezoelectric softening, when passing from the
FE to the AFE phase (note 12 of Ref. \citenum{CCT16c}). We believe that, in
the case of MAPbI$_3$, the stiffening when entering in the O phase has two
main causes: the loss of mobility of the MA molecules and the stabilization
of shorter and hence stiffer N$-$H$\cdot \cdot \cdot $I bonds.\cite{LBL16}
As shown by high frequency dielectric,\cite{PW87} inelastic neutron
scattering,\cite{CFI15} NMR\cite{BWR18} and ultrafast 2D vibrational
spectroscopy\cite{BSB15} experiments, the mean reorientation time of MA with
electric dipole $\mu $ in the C and T phases is extremely short, $\tau \leq
10^{-10}$~s, and causes a Debye dielectric relaxation\cite{PW87}
\begin{equation}
\epsilon =\epsilon _{\infty }+\frac{C}{T}\frac{1}{1-i\omega \tau }
\label{Debye}
\end{equation}%
\begin{equation}
C=\frac{N\mu ^{2}\eta }{3k_{\text{B}}\epsilon _{0}},\;\eta =\frac{2+\epsilon
_{0}}{3}  \label{Curie}
\end{equation}%
where $\epsilon _{0}$ is the vacuum permittivity, $k_{\text{B}}$ the
Boltzmann constant, $N=$ $4.0\times 10^{27}$~m$^{-3}$ the number of MA per
unit volume, $\tau =\tau _{0}\exp \left( W/T\right) $ their relaxation time,
close to the reorientation time. The MA posses also an elastic dipole\cite%
{NH65,NB72} with anisotropic component $\Delta \lambda $, and therefore the
compliance $s=E^{-1}$, which is the mechanical analogue of the dielectric
susceptibility $\chi \simeq \epsilon $, must contain a contribution
identical to Eq. (\ref{Debye}) but with $\Delta s\propto \left( \Delta
\lambda \right) ^{2}$ instead of $C$. Then, there is a maximum in the
losses at $\omega \tau =1$, while for shorter $\tau $,
or higher $T$, the mechanical loss is negligible and the compliance softened
of $\Delta s$. Extrapolating the Arrhenius law $\tau =\tau
_{0}\exp \left( W/T\right) $ with $\tau _{0}=5.4\times 10^{-12}$~s and $%
W=910 $~K found from high frequency dielectric relaxation,\cite{PW87} we
find that the condition $\omega \tau =1$ would occur around 50~K at our
frequencies, but, if the MA get abruptly blocked in the O phase, then the
restiffening to the unrelaxed compliance occurs sharply at $T_{\mathrm{OT}}$%
, contributing to the observed anomaly. Further stiffening below $T_{\mathrm{%
OT}}$ may result from the strengthening of the shorter H$\cdot \cdot \cdot $%
I bonds. The loss of mobility of the MA must not be complete at $T_{\mathrm{%
OT}}$, however, because still an important anelastic relaxation, labeled AR
in Fig. \ref{fig-anel}, is observed within the O phase.

\begin{figure}[tbh]
\includegraphics[width=8.25 cm]{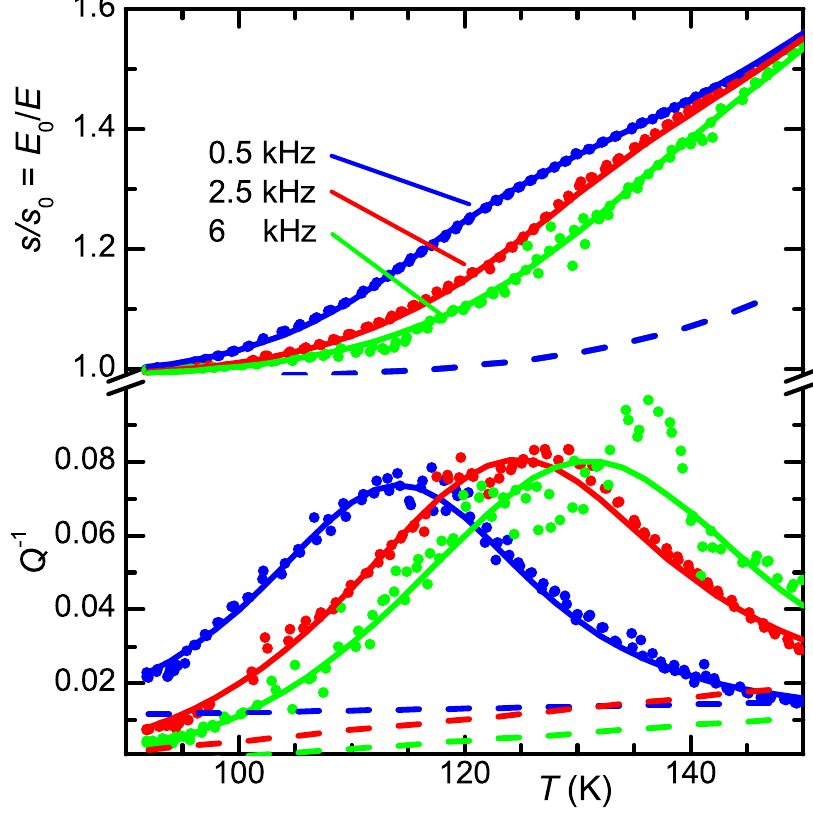}
\caption{Fit with Eq. (\ref{s-rel}) to the real part (upper panel) and
losses (lower panel) of the dynamic compliance of MAPbI$_{3}$.}
\label{fig-fit-anel}
\end{figure}

This relaxation is much broader than simple Debye, and the fit in Fig. \ref%
{fig-fit-anel} has been obtained with
\begin{equation}
s=s_{bg}+\frac{\Delta s}{\left[ 1+\left( i\omega \overline{\tau }\right)
^{-\alpha }\right] ^{\gamma }}  \label{s-rel}
\end{equation}%
\[
\overline{\tau }^{-1}=\tau ^{-1}\cosh \left( A/2T\right)
\]%
\[
\Delta s=\frac{\Delta }{T\cosh ^{2}\left( A/2T\right) }
\]%
which is the Havriliak-Negami expression \bigskip for broadened dielectric
relaxations, with the inversion of the Cole-Davidson distribution with respect
to small and large $\tau $,\cite{Fan58} which follows the Arrhenius law.
An average asymmetry $A$ between the initial and final energies of the
elementary relaxation events,\cite{Cor93} has been introduced in order to reproduce the
increase of the intensity with $T$, instead of the decrease as $1/T$.
For the background (dashed lines)\ contribution to the real
part we chose $s_{bg}=$ $s_{0}+a/\left[ \exp \left( b/T\right) -1\right] $,
and linear backgrounds for the losses. The main fitting parameters are: $%
\tau _{0}=2.0\times 10^{-12}$~s, $W=2340$~K, $\alpha =0.797$, $\gamma =0.498$%
, $A=366$~K. It results that the relaxation times not only are broadly
distributed ($\alpha ,\gamma <1$), but also highly correlated. In fact, $%
\tau _{0}$ is two orders of magnitude larger than for typical independent
relaxations of molecules and point defect, while it is characteristic of
extended/correlated defects, such as domain walls and dislocations.\cite%
{NB72}

\begin{figure}[tbh]
\includegraphics[width=8.25 cm]{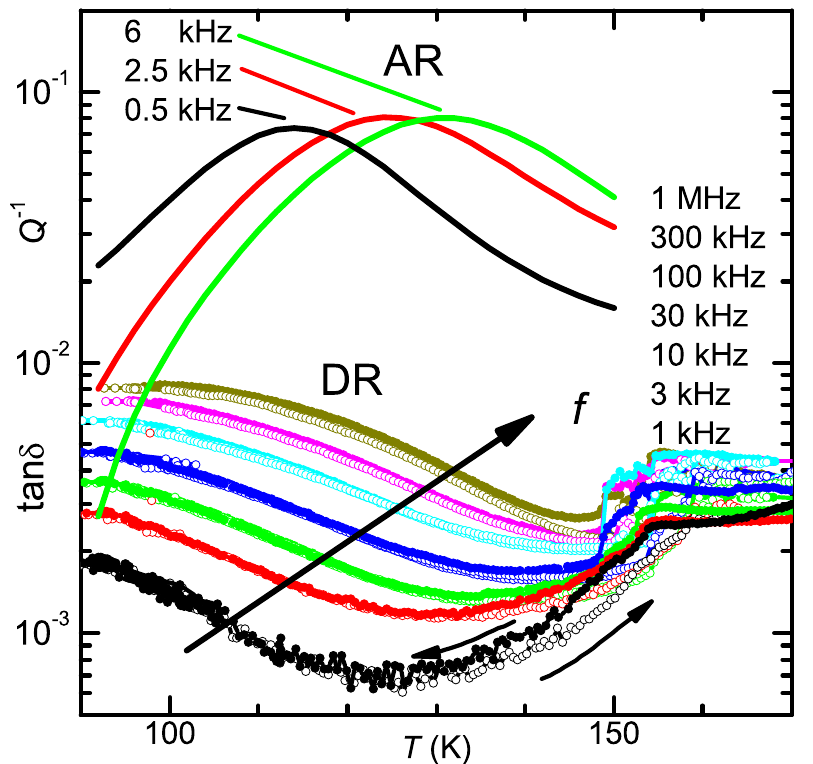}
\caption{Comparison between the dielectric and anelastic losses measured at several
frequencies during cooling and heating on the same sample of Fig. \protect\ref{fig-fit-anel}.
For clarity, only the anelastic fitting curves are shown.}
\label{fig-diel-rel}
\end{figure}

Indeed, relaxation of the walls between the different variants of O domains
is expected, and the same is true within the T phase, where the $%
Q^{-1}\left( T\right) $ curves are much broader and with little dispersion
in frequency, so conforming to the typical signature of domain wall
relaxation. The fitted peak in the O phase, instead, is much more sharp and
intense than the DW relaxation within the T phase, and this suggests that a
considerable fraction of MA contributes to it. Also in the paraelectric
phase of [(CH$_{3}$)$_{2}$NH$_{2}$]Co(HCOO)$_{3}$ an anelastic relaxation
process has been found by Resonant Ultrasound Spectroscopy\cite{TJC12} (RUS) and
attributed to the reorientation of the dimethylammonia. In that case, in the
absence of measurements at several frequencies, it has been assumed that the
process is simple Debye. In our case, instead, we can also say that the
strong correlations between neighboring MA, mediated by octahedral tilting,
must be AFE, because the dielectric relaxation does not show the same process
as the anelastic one (see Fig. \ref{fig-diel-rel} later on).
Indeed, the MA in the $Pnma$ phase are AFE ordered along the $b$ axis.\cite{SHS03,WHG16}
Therefore, the anelastic relaxation may arise, for example, from clusters
of an even number of MA, which reorient by $\sim 90^{\circ}$
together with the surrounding octahedral distortions, maintaining the AFE
correlation, and hence without a net change of the electric polarization.

Before presenting the dielectric data, we mention that the mechanical
properties of MAPbI$_3$ have recently been measured also by RUS\cite{HPS18}
and ultrasound propagation on crystals,\cite{ABG17b} probing higher frequencies and
presumably elastic constants little sensitive to the MA reorientations,
which occur mainly in the tetragonal $ab$ plane. Those measurements cannot be
directly compared with the Young's modulus reported here, which,
being a polycrystalline average, contains the contributions of
all the elastic constants.

Figure \ref{fig-diel-rel} shows the complex permittivity measured on a piece
of the sample used in Fig. \ref{fig-fit-anel}, together with the fitting
anelastic curves, and it is clear that a dielectric relaxation exists, but
much slower than the anelastic one and with small intensity. Being a very
broad relaxation, it might be related to domain walls.
This dielectric relaxation, though with smaller intensity, is compatible
with the high temperature tails of previous measurements extended to lower
temperatures\cite{OMS92,FHE16} (see Fig. \ref{fig-S4}).

\begin{figure}[tbh]
\includegraphics[width=8.5 cm]{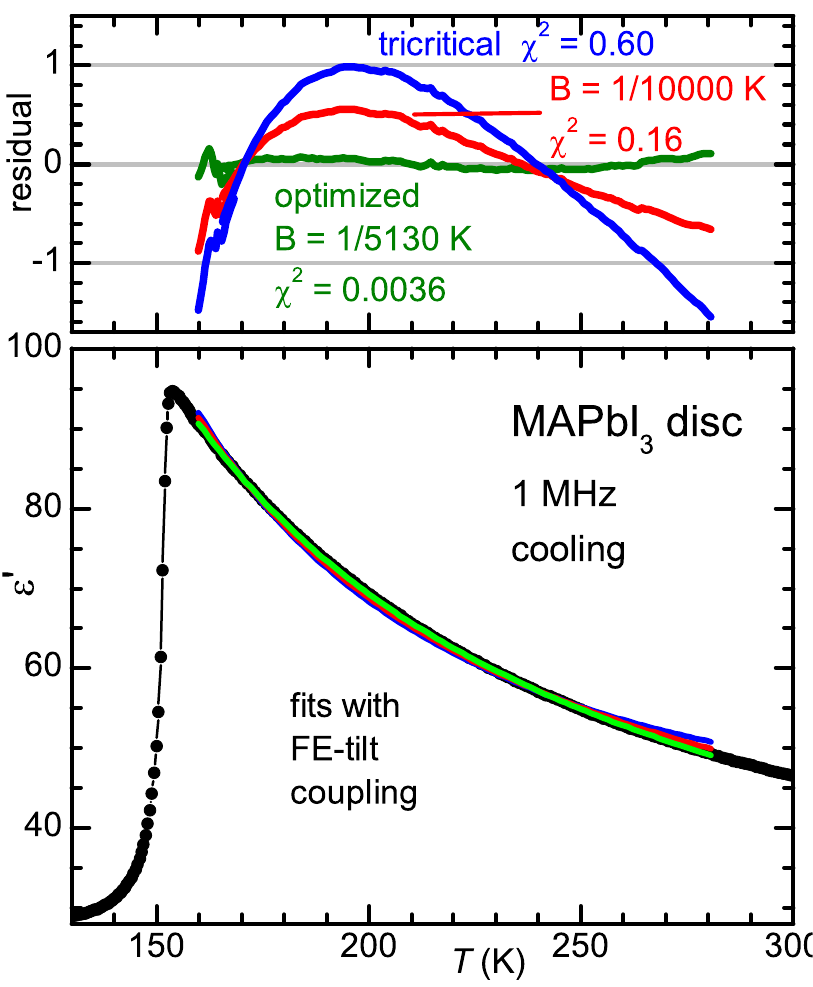}
\caption{Real part of the permittivity measured during cooling on a disc of
MAPbI$_{3}$ at 1~MHz. For the full measurement at other frequencies see SI.
The three fitting curves are explained in the text; their residuals are
shown in the upper panel.}
\label{fig-fit-diel}
\end{figure}

We pass now to the real part of the dielectric permittivity, measured on
discs at frequencies up to 1~MHz (see Fig. \ref{fig-fit-diel} and Fig. \ref{fig-S3}),
which substantially confirms previous measurements:\cite{OMS92,Ges97,GKB17,ABG17b}
no sign of a transition justifying a normal FE\ state at room temperature,
but an increase of the Curie-Weiss type up to $T_{\mathrm{OT}}$,
which suggests that, in the absence of the O/T tilt
transition, a FE state might be possible below that temperature. As already
mentioned, the permittivity at 90~GHz\cite{PW87} has been fitted with a
Debye relaxation, Eqs. (\ref{Debye}-\ref{Curie}), finding the Arrhenius
parameters of $\tau $ and a dipole moment $\mu =0.85$~D for MA in MAPbI$_3$
(1~Debye = $3.3\times 10^{-30}$~C$~$m, but using the data in Table II of
Ref. \citenum{PW87} we find $\mu =2.8$~D for MAPbI$_3$), not far from
other estimates of $\mu \simeq 1-3$~D from low frequency dielectric and first principle
calculations.\cite{OMS92,GKB17,FBB14}

The extrapolation of the Arrhenius law so found for $\tau $ to our maximum
frequency of 1~MHz, shifts the region of frequency dispersion, $\omega \tau
\sim 1$, below 90~K, and any frequency dispersion in the T and C phases
comes from much slower processes, \textit{e.g.} Maxwell-Wagner from free
charge carriers. These relaxations are present also in our samples above
room temperature (Fig \ref{fig-S3}), but negligible at 1~MHz below 280~K. In these
conditions, it is possible to probe the MA dipoles through the $C/T$ term in
Eq. (\ref{Debye}), which becomes $\sim C/\left( T-T_{\mathrm{C}}\right) $ if
the MA cooperatively order themselves in a FE phase below $T_{\mathrm{C}}$.
In the context of molecular reorientation, the Clausius-Mosotti formula is
used, which can be rewritten as (see SI)
\begin{eqnarray}
\epsilon &=&\epsilon _{\infty }+\frac{\left( \epsilon _{\infty }+2\right) T_{%
\mathrm{C}}}{T-T_{\mathrm{C}}}  \label{LL} \\
T_{\mathrm{C}} &=&\frac{\left( \epsilon _{\infty }+2\right) N\mu ^{2}}{%
27\epsilon _{0}k_{\text{B}}}  \label{TC-mu}
\end{eqnarray}%
namely the Curie-Weiss formula with Curie constant
\begin{equation}
C=\left( \epsilon _{\infty }+2\right) T_{\mathrm{C}}~.  \label{C-TC}
\end{equation}%
Attempts to fit our data with Eq. (\ref{LL}) fail, because of the excessive
curvature of the fitting curves. The permittivity is calculated
selfconsistently,\cite{Hip54,Raj03} taking into account that the local field
felt by the dipole is enhanced by the polarization of the surrounding
dipoles. Then, the situation is improved using models where the local field
is decreased with respect to the Lorentz field,\cite{Ons1936,Hip54,Raj03}
and so is also the rise of $\epsilon $ with respect to $1/T$, due to the
cooperative ordering. Accordingly, $\epsilon \left( T\right) $ of MAPbI$_3$
has been fitted with the formula of Onsager,\cite{Ons1936,Hip54,Raj03} but
since its curvature is now too low to fit the data, the term $1/T$ has been
substituted with $1/\left( T-T_{\mathrm{C}}\right) $.\cite{OMS92,CTB16} The
fits obtainable in this manner are good, and this is true also for our data,
but the problem is that the introduction of $T_{\mathrm{C}}$ in the Onsager
formula does not correspond to any physical model. In Onsager's model the
effect of a dipole on its neighbors is totally eliminated, so eliminating
from the start the possibility of FE ordering, and the significance
of the introduction of $T_{\mathrm{C}}$ and of the magnitude of the MA
dipole extracted in this manner is dubious.


We discuss therefore another possible origin of the peculiar rise of $%
\epsilon \left( T\right) $ below room temperature:
the coupling between the reorienting MA and the tilting of the surrounding
PbI$_{6}$ octahedra. This interaction should not be ignored, especially in
view of the recent suggestion that the tilt transitions are due more to the
H bonds between MA and the halides than to the shrinking of the MA$-$I
network with respect to the less compressible PbI$_{6}$,\cite{LBL16} as
usual in other perovskites.\cite{CCT14c}
\begin{figure}[tbh]
\includegraphics[width=8.25 cm]{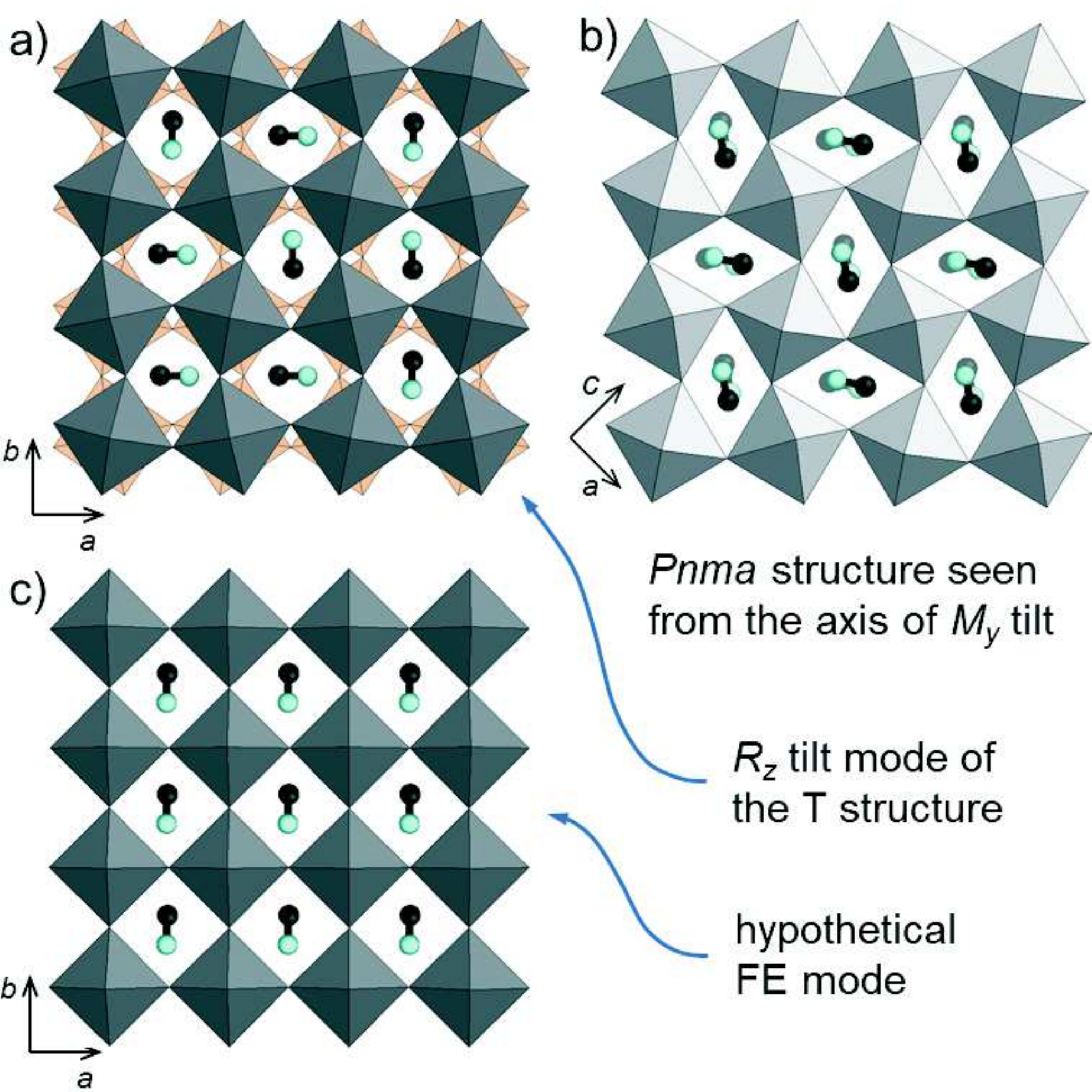}
\caption{a) $R_{z}$ tilt mode $a^{0}a^{0}b^{-}$, producing the T phase, seen
from the $c$ axis, with the successive layer of octahedra tilted
out-of-phase; it corresponds to the order parameter $Q$
in Eq. (\ref{Landau}); the MA are represented without the H atoms and
disordered within the $ab$ plane. b) Two layers of octahedra and
MA molecules in the $Pnma$ structure,\protect\cite{WHG16}
seen from the $b$ axis of the $M$ mode (notice that in the $Pnma$ setting the
tetragonal $c$ axis becomes $b$); the MA molecules have AFD correlation
within the $ac$ plane and AFE along $b$. c) a
 ferroelectric mode
with parallel MA molecules, corresponding to $P$ in Eq. (\ref{Landau}).}
\label{fig-tilts}
\end{figure}

In-phase tilting of the octahedra along an axis creates an
antiferrodistortive (AFD) pattern, that competes with the FE ordering of the MA
dipoles, as shown in Fig. \ref{fig-tilts}b). This is the case of the O $Pnma$
ground state of MAPbI$_3$,\cite{WHG16} with tilt pattern $a^{-}b^{+}a^{-}\ $
in Glazer's notation (SI), where in-phase tilting about the $b$ axis (called
$M_{y}$ mode) creates rhombic channels, and the MA molecules lay within the $ac$ plane,
arranged approximately according to the longer diagonals of the rhombi and
in AFE manner along the $b$ axis.\cite{WHG16}
Though $a^{-}b^{+}a^{-}$ tilting is a rather
common ground state of perovskites with strong tendency to tilting, in the
case of MAPbX$_{3}$ it seems to be favored by the formation of short N$-$H$%
\cdot \cdot \cdot $X hydrogen bonds, and probing the instantaneous local
structure suggests that this type of tilts and distortions persists even in
the C phase.\cite{BSS16,BM17} It can be concluded that coupling
between the tilt $M$ and the FE modes, where the MA must be all parallel
(Fig. \ref{fig-tilts}c)), is competitive. The situation is open for the $R$
modes, in antiphase along the tilt axis (Fig. \ref{fig-tilts}a); in that
case, the MA molecule sits in the middle of two rhombic spaces at 90$^{\circ}$
with each other, so that, in the absence of additional interactions, there
is no need for the MA molecules to arrange themselves in an AFD manner. This
is confirmed by PZT, a classic FE: in its rhombohedral FE phase it undergoes
$a^{-}a^{-}a^{-}$ tilting, and the concomitant small increase of $\epsilon $
and of the polarization indicate slightly cooperative coupling between $%
a^{-}a^{-}a^{-}$ and FE modes.\cite{CCT14c}

The effect of coupling between two modes can be described in a
phenomenological manner within the Landau theory of phase transitions. The
simplest possible expansion of the free energy for our purpose is (SI and
Ref. \citenum{BT93})

\begin{equation}
F=\frac{\alpha _{2}}{2}P^{2}+\frac{\alpha _{4}}{4}P^{4}+\frac{\beta _{2}}{2}%
Q^{2}+\frac{\beta _{4}}{4}Q^{4}+\frac{\gamma }{2}P^{2}Q^{2}  \label{Landau}
\end{equation}%
where the two order parameters are the polarization $P$ and tilt angle $Q$,
and $\alpha _{2}=\alpha _{0}\left( T-T_{\mathrm{C}}\right) $, $\beta
_{2}=\beta _{0}\left( T-T_{\mathrm{T}}\right) $ induce FE and tilt
transitions below $T_{\mathrm{C}}$ and $T_{\mathrm{T}}$, in the absence of
coupling ($\gamma =0$). The biquadratic coupling term is the lowest order
mixed term permissible by symmetry in this case. As shown in SI, the
dielectric susceptibility obtained from Eq. (\ref{Landau}) in the
paraelectric and untilted phase is simply $\chi = \alpha _{0}^{-1}/\left(
T-T_{\mathrm{C}}\right) ,$ \textit{i.e.} the Curie-Weiss law with Curie
constant $C=\alpha _{0}^{-1}$, unaffected by tilting. This means that, in
the temperature range of interest $T_{\mathrm{OT}}<T<$ $T_{\mathrm{TC}}$, we
can ignore the lower tilt transition at $T_{\mathrm{OT}}$, identify $Q$ with
the $R_{z}$ mode condensed in the T phase, and set $T_{\mathrm{T}}=T_{%
\mathrm{TC}}$ and $T_{\mathrm{C}}\leq T_{\mathrm{TC}}$, since no trace of FE
transition is found in the dielectric susceptibility and compliance above $%
T_{\mathrm{TC}}$. In the T phase, the permittivity becomes (SI)
\begin{eqnarray}
\epsilon &=&1+\frac{C}{T-T_{\mathrm{C}}+B~C\left( T_{\mathrm{T}}-T\right) }
\label{tilt0} \\
C &=&1/\alpha _{0}~,\;B=\gamma \beta _{0}/\beta _{4}  \nonumber
\end{eqnarray}%
This formula is the Curie-Weiss law with a correction term $BC\left( T_{%
\mathrm{T}}-T\right) $ to $T-T_{\mathrm{C}}$. Besides changing the curvature
of $\epsilon \left( T\right) $, this shifts the onset of the FE transition,
when $\chi $ diverges, to
\[
T_{\mathrm{FE}}=\frac{T_{\mathrm{C}}-B~C~T_{\mathrm{T}}}{1+B~C}
\]%
so that a cooperative (competitive) interaction with the tilt mode raises
(lowers) $T_{\mathrm{FE}}$ with respect to $T_{\mathrm{C}}$. The above
formula assumes $\epsilon _{\infty }=1$; in order to include $\epsilon
_{\infty }$, and to relate the phenomenological constants to the magnitude $%
\mu $ of the electric dipole of the reorienting MA, we compare Eq. (\ref%
{tilt0}) with Eqs. (\ref{LL}$-$\ref{TC-mu}) and insert there the correction
for polar-tilt coupling:
\begin{equation}
\epsilon =\epsilon _{\infty }+\frac{C}{T-T_{\mathrm{C}}+B~C\left( T_{\mathrm{%
T}}-T\right) }  \label{chi-tilt}
\end{equation}

Figure \ref{fig-fit-diel} shows the best fit (green curve)\ with Eq. (\ref%
{chi-tilt}) of $\epsilon \left( T\right) $ measured in a disc at 1~MHz
during cooling. The quality of the fit is the same as with formulae like
that of Onsager, but
the values and meanings of the parameters are quite different. The best fit (%
$\chi ^{2}=0.0036$) with $T_{\mathrm{T}}$ fixed at $328~$K, yields $B=1/5134$%
~K, $T_{\mathrm{C}}=223.6$~K, $\epsilon _{\infty }=12.26$, from which it
results $C=3190$~K, $\mu =1.07$~D and $T_{\mathrm{FE}}=-25$~K. Therefore,
the coupling between the FE and tilt modes is competitive and the FE
transition is even suppressed at any temperature. The fit is quite sensitive
to the magnitude of this coupling, as demonstrated by the fact that doubling
$B$ (red curve) increases $\chi ^{2}$ of nearly 50 times. The origin for
such a strong competitive coupling between the $R$ tilt mode and the FE MA
ordering may be that the tendency to form short N$-$H$\cdot \cdot \cdot $I
bonds leading to the distortions of the $Pnma$ ground state exists also in
the T phase.\cite{BSS16,BM17} The tilt transition at $T_{\mathrm{T}%
}=T_{\mathrm{TC}}=328$~K has been reported to be first-order, as we
find here, but also close to tricritical.\cite{WHG16} We tried to fit also
with the tricritical free energy, by substituting the term $\frac{b}{4}%
Q^{4}$ with $\frac{b}{6}Q^{6}$, but the fit yielded similar
values of the various parameters with a definitely worse quality (blue curve
with $\chi ^{2}$ 165 times larger than the best fit). The
description of first order transitions requires the inclusion of both $%
Q^{4}$ and $Q^{6}$ terms, which may somewhat improve the
fit, thanks to the additional parameter, but with scarce additional insight,
in view of the excellent result already obtainable with the simplest free
energy.

In conclusion, we presented anelastic ($0.5-6$~kHz) and dielectric (1~kHz$-1$~MHz)
measurements on MAPbI$_3$, which can be fitted very well within
a comprehensive picture of the reorientation dynamics of the electric
and elastic dipoles associated with the MA molecules, their correlations
and their coupling with the tilt modes of the PbI$_3$ octahedra.
It results that the coupling with the ferroelectric and antiferrodistortive
modes is competitive, and prevents the ordering into a ferroelectric phase.
At the onset of the tilt transition to the orthorhombic structure
the reorientation dynamics
of the molecules is strongly hindered, resulting in stiffening of both the
dielectric and elastic susceptibilities. Yet, a considerable fraction of
MA can still reorient in the orthorhombic phase as clusters of
antiferroelectrically correlated dipoles, which give rise to intense
anelastic relaxation without a dielectric counterpart.

\begin{acknowledgement}

The authors thank Paolo Massimiliano Latino (ISM-Tor Vergata) for
his technical assistance, Sara Notarantonio (ISM-Montelibretti) for
the assistance in the synthesis and Dr. Gloria Zanotti for fruitful
discussions.

\end{acknowledgement}

\section{Supplementary Information}

\subsection{Experimental methods}

\subsubsection{Synthesis}

All starting chemicals were purchased from Sigma-Aldrich; methylammonium
iodide (CH$_{3}$NH$_{3}$I) was synthesized following the method reported in
Ref. \citenum{SMP15}: hydroiodic acid (HI) (aqueous solution 40\%) was added
stoichiometrically dropwise to methylamine (CH$_{3}$NH$_{3}$) (aqueous
solution 57\%) under stirring at $0^{\circ}$C, and the solution was left 2~h under stirring, then the solvent was
removed by rotary evaporation and the solid was washed with diethylether
several times and dried under vacuum to yield a white crystalline solid.

MAPbI$_3$ (CH$_{3}$NH$_{3}$PbI$_{3}$) was synthesized following the procedure
reported in Ref. \citenum{JHZ17} with slight modifications;
2.80~g of PbI$_2$ and 2.90~g of CH$_3$NH$_3$I were mixed and ground in a mortar for
at least 5', then the mixture was added to 100~ml of glacial acetic acid in
a flask placed in an ultrasonic bath at ambient temperature in agitation for
20'; the solid was decanted and after 8~h was filtered and repeatedly washed
with de-aerated anhydrous ethanol and finally with n-hexane. The glossy
black microcrystalline solid was then dried under vacuum at $60^{\circ}$C.

\subsubsection{XRD}

\begin{figure}[h]
\includegraphics[width=14 cm]{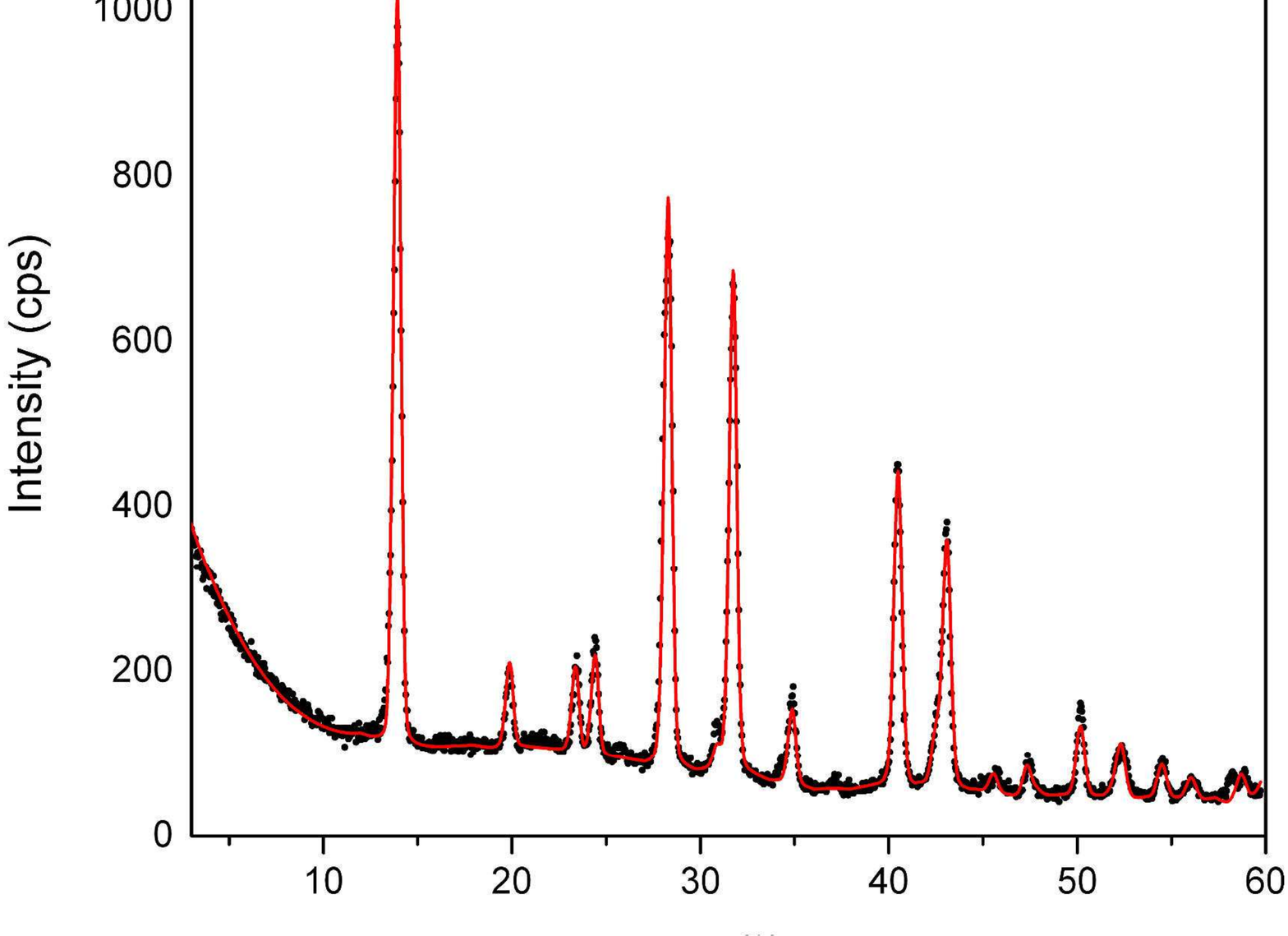}
\caption{XRD pattern of MAPbI$_3$ sample (black dots) with Rietveld
fit (red line)}
\label{XRD}
\end{figure}

X-ray diffraction (XRD) measurements were performed on a Seifert TT 3003
diffractometer equipped with a secondary graphite monochromator, using
Cu K$\alpha$ radiation ($\lambda=$ 1.5418 \AA ). Data were collected in the
Bragg-Brentano geometry, in the $2\theta$ range $3-60$ degrees and
counting 4~s/step. The powder sample was protected from air humidity with
a Mylar cap, to avoid its gradual decomposition at room temperature.

The XRD pattern of a MAPbI$_3$ sample is shown in Fig. S\ref{XRD}. All peaks were indexed
according to the tetragonal $I4/mcm$ phase. No degradation, with the formation
of PbI$_2$ and/or MAPbI$_3$ hydrates, was observed. From the Rietveld refinement\cite{You93}
of data, the MAPbI$_3$ lattice parameters $a=$ $b=$ 8.888(2)~\AA\ and $c=$
12.616(6)~\AA\ were obtained, in agreement with the data reported in the
literature.\cite{BFK13}

\subsubsection{Pressing}

The powder was pressed at 9~tons in discs of 13~mm of diameter and at
10~tons at $\sim 70^{\circ}$C in bars $40\times 6\times 0.4-0.7$~mm$^{3}$.
The powders and samples were uniformly black and were
exposed to air only for the short times necessary to press them, apply
electrodes with Ag paint, and mount them in the air-tight setups for
anelastic and dielectric measurements.

\subsection{Anelasticity and anelastic measurements}

The term "anelastic" was introduced by Zener\cite{Zen48} to indicate the
time/frequency dependent strain response to an applied stress, in addition
to the instantaneous elastic response. The frequency dependent mechanical
susceptibility, the compliance $s=s^{\prime }-is^{\prime \prime }$,
acquires an imaginary part describing the retarded response and the
consequent mechanical losses.
The anelastic response is due to some internal variable coupled to strain
and having its own dynamics, usually expressed in terms of a relaxation
time $\tau$ or a spectrum of times.
Examples are: point or extended defects,
that may change between at least two states with different local strain
(elastic dipole); heat flow coupled with change of volume; polarons.
The anelastic and dielectric responses may be described with the same
formulas, using the correspondence stress $\leftrightarrow$ electric field,
strain $\leftrightarrow$ electric polarization, electric dipole
$\leftrightarrow$ elastic dipole. The theory of anelastic and
dielectric relaxation from point defects
has been systematized by Nowick and Heller.\cite{NH65,NB72}. A major
difference between the two phenomena is that the elastic and anelastic
response is insensitive to free charges, making it an effective tool
for probing defects and phase transitions also in metals and highly
conducting materials.

\begin{figure}[h]
\includegraphics[width=8.5 cm]{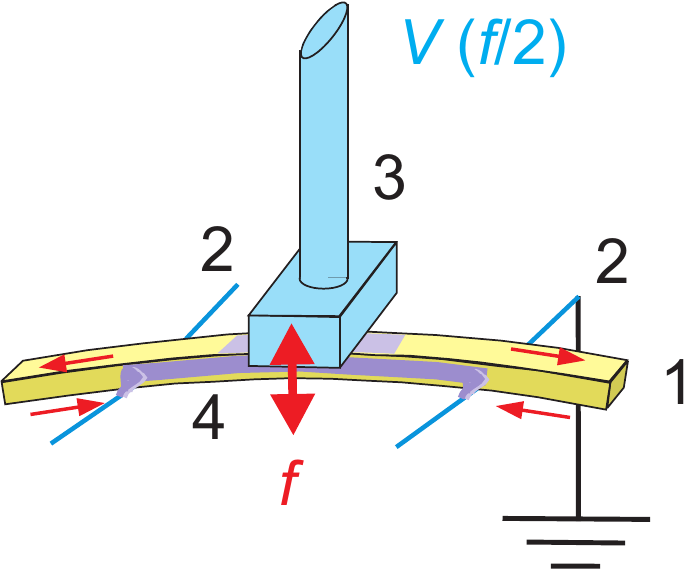}
\caption{Schematization of the thin bar sample (1) suspended on two
thermocouple wires (2) and electrostatically excited in flexure by the
electrode (3). The gray color (4) indicates Ag paint used to fix the
sample to the wires, short them and act as counterelectrode.}
\label{flex}
\end{figure}

The complex Young's modulus $E=E^{\prime }+iE^{\prime \prime }$, or its
reciprocal, the compliance $s=s^{\prime }-is^{\prime \prime }$, was measured
by suspending the bar on thin thermocouple wires in high vacuum ($\sim
1\times 10^{-6}$~mbar or up to 0.2~mbar He) and electrostatically exciting
their odd flexural modes. The variable temperature insert and the sample holder
are described in detail in Ref. \citenum{CDC09}. Figure S\ref{flex} schematically
shows the sample on the wires and the electrode. The application of a
sinusoidal voltage $V\left( t\right) = V_0 sin\left(\omega t/2 \right)$ induces an opposite
voltage on the top of the sample, which is made conductive with Ag paint and
grounded through one of the two thermocouple wires. The resulting electrostatic
attraction is $\propto V^2 \propto sin\left(\omega t \right)$.
The fact that the sample vibrates at twice the excitation frequency allows
the same electrode to be used for the excitation and the detection of
the vibration through the modulation of the sample/electrode capacitance.
Such a capacitance is part of an external high frequency ($\sim 12$~MHz)
oscillator, whose modulated frequency is demodulated and provides the vibration
amplitude of the sample.
The red arrows in Fig. S\ref{flex} show that during flexure one side of the
sample undergoes extension and the other compression, so that what is
probed is the Young's modulus.
The resonance frequencies are $f_{n}\propto \alpha _{n}\sqrt{E^{\prime }}$,%
\cite{NB72} where $\alpha _{n}$
is the geometrical factor for the $n$-th mode, and therefore the Young's
modulus normalized to a reference temperature $T_{0}$ can be plotted as $%
E\left( T\right) /E\left( T_{0}\right) =$ $f^{2}\left( T\right) /f^{2}\left(
T_{0}\right) $. Up to three modes could be measured during a same run. The
elastic energy loss $Q^{-1}=$ $E^{\prime \prime }/E^{\prime }=$ $s^{\prime
\prime }/s^{\prime }$ was measured from the free decay or from the resonance
curve under forced vibration.

\subsection{Dielectric measurements}

The complex dielectric permittivity, $\epsilon =$ $\epsilon ^{\prime
}-i\epsilon ^{\prime \prime }$, with losses $\tan \delta =\epsilon ^{\prime
\prime }/\epsilon ^{\prime }$, was measured with a HP 4284A LCR meter with a
four-wire probe and an electric field of 0.5 V/mm up to 1~MHz.
Temperature was controlled with a modified Linkam HFS600E-PB4 stage filled with N$_{2}$.

\newpage

\subsection{Additional dielectric curves}

\begin{figure}[h]
\includegraphics[width=12 cm]{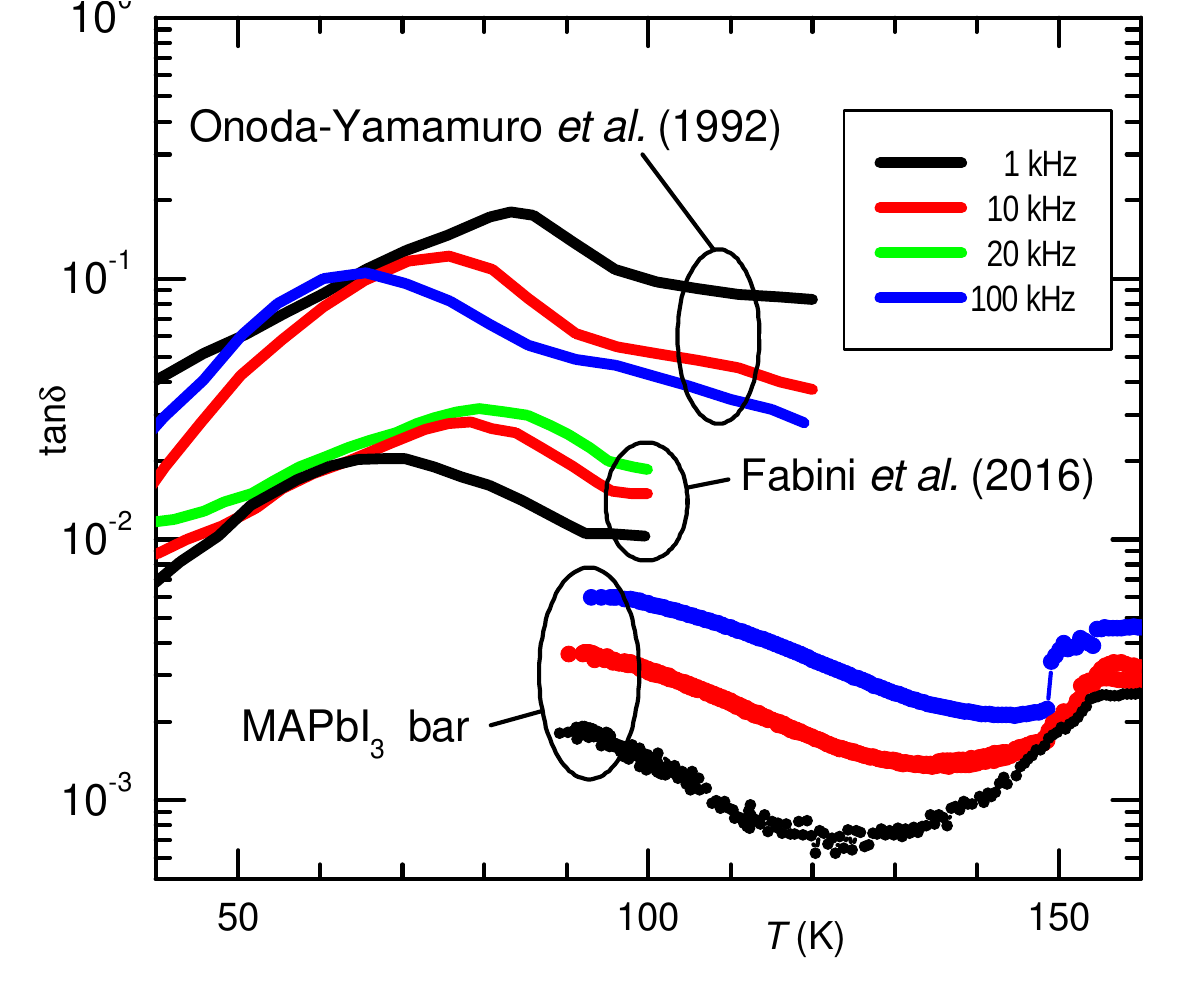}
\caption{Comparison between some of the $%
\tan \protect\delta $ curves of Fig. 3, measured on the bar for the
anelastic experiments, and those of Refs. \protect\citenum{OMS92} and
\protect\citenum{FHE16} at the same frequencies.}
\label{fig-S4}
\end{figure}

The dielectric relaxation we measure in the O phase, though with smaller
intensity, is compatible
with the high temperature tails of previous measurements extended to lower
temperatures\cite{OMS92,FHE16} (Fig. \ref{fig-S4})). The differences in amplitude
between these measurements may arise from the fact that the residual
molecular motions probed at such low temperatures strongly depend on
extrinsic factors, such as defects, and the precise domain wall
configurations, and the latter depend also on the sample thermal history.

\newpage

\subsection{Glazer's notation for the octahedral tilt patterns}

The notation of Glazer\cite{Gla72} describes the rotations of the octahedra in a
cubic perovskite: $a^{+}$ stands for tilting of all the
octahedra along an axis of the same angle $a$; $a^{-}$ for antiphase
tilting, \textit{i.e.} alternately $+a$ and $-a$ along the axis; $a^{0}$ for
no rotation. The rotations about the $x,$ $y$ and $z$ pseudocubic axes are written in
sequence.

\subsection{Permittivity with polarization-tilt coupling from Landau's theory}

The simplest free energy describing a tilt and a FE transition with coupling
is (see \textit{e.g.} Ref. \citenum{BT93})

\begin{eqnarray}
G &=&\frac{\alpha _{2}}{2}P^{2}+\frac{\alpha _{4}}{4}P^{4}-EP+\frac{\beta
_{2}}{2}Q^{2}+\frac{\beta _{4}}{4}Q^{4}+\frac{\gamma }{2}P^{2}Q^{2}
\label{Landau} \\
\alpha _{2} &=&\alpha _{0}\left( T-T_{\mathrm{C}}\right) \\
\beta _{2} &=&\beta _{0}\left( T-T_{\mathrm{T}}\right)
\end{eqnarray}%
where $\alpha _{0},\beta _{0},\alpha _{4},\beta _{4}>0$ and the coefficients
of the quadratic terms decrease linearly with temperature, so that above $T_{%
\mathrm{C}}$ and $T_{\mathrm{T}}$ the minimum of $G$ is for $P=Q=0$
(symmetric phase), but below those temperatures double-well minima develop
at finite values of $P=P_{0}\left( T\right) $ and $Q=Q_{0}\left( T\right) ,$
the spontaneous polarization and tilt angle in the FE and tilted phases. The
coupling coefficient $\gamma $ may be positive (competitive) or negative
(cooperative), and coupling terms linear in either $P$ or $Q$ are forbidden
by symmetry in the high temperature symmetric phase, whose energy is
invariant for changes of the sign of either order parameters. The $-EP$ term is the
Legendre transformation $G=F-EP$ from the Helmholtz free energy $F$, whose
differential is $dF=-SdT+EdD$, to the Gibbs free energy, whose differential
is $dG=-SdT-DdE$; in this manner the independent variable changes from $D/P$
to $E$ and we can calculate the susceptibility $\chi =dP/dE$ ($D=\epsilon
_{0}E+P\cong P$ because the external field is null, except for the small
alternate probe field to measure $\epsilon =1+\chi $).

The equilibrium values of the order parameters in the minima of $G$ are found from

\begin{eqnarray}
0 &=&\frac{\partial G}{\partial P}=\alpha _{2}P+\alpha _{4}P^{3}+\gamma
PQ^{2}-E  \label{eq-P} \\
0 &=&\frac{\partial G}{\partial Q}=\beta _{2}Q+\beta _{4}Q^{3}+\gamma P^{2}Q
\label{eq-Q}
\end{eqnarray}

We need to calculate $\chi =dP/dE$ and also $dQ/dE$, which can be obtained
in implicit form differentiating the two above equations:%
\[
0=\frac{d}{dE}\frac{\partial G}{\partial Q}\rightarrow \frac{dQ}{dE}=-\frac{%
2\gamma PQ}{\beta _{2}+3\beta _{4}Q^{2}+\gamma P^{2}}\frac{dP}{dE}
\]

\begin{eqnarray}
0 &=&\frac{d}{dE}\frac{\partial G}{\partial P}\rightarrow 1=\frac{dP}{dE}%
\left( \alpha _{2}+3\alpha _{4}P^{2}+\gamma Q^{2}\right) +2\gamma PQ\frac{dQ%
}{dE}\rightarrow  \nonumber \\
\chi &=&\frac{dP}{dE}=\left[ \alpha _{2}+3\alpha _{4}P^{2}+\gamma Q^{2}-%
\frac{4\gamma ^{2}P^{2}Q^{2}}{\beta _{2}+3\beta _{4}Q^{2}+\gamma P^{2}}%
\right] ^{-1}=  \label{chi-tilt} \\
&=&\frac{a+3bQ^{2}+\gamma P^{2}}{\left( \alpha +3\beta P^{2}+\gamma
Q^{2}\right) \left( a+3bQ^{2}+\gamma P^{2}\right) -4\gamma ^{2}P^{2}Q^{2}}
\end{eqnarray}

For $T>T_{\mathrm{T}},T_{\mathrm{C}}$ it is $P=Q=0$, and
\begin{equation}
\chi =\frac{1}{\alpha _{0}\left( T-T_{\mathrm{C}}\right) }  \label{chi0}
\end{equation}%
Therefore, the dielectric susceptibility is the Curie-Weiss law with Curie
constant $C=1/\alpha _{0}$, unaffected by the coupling with tilting above
the tilting transition, at least in the present approximation of neglecting
fluctuations above the transition temperatures. Then, in the temperature
range of interest, $T_{\mathrm{OT}}<T<$ $T_{\mathrm{TC}}$, we can ignore the
second tilt transition at $T_{\mathrm{OT}}$ and identify $Q$ with the $R_{z}$
mode condensed in the T phase, and set $T_{\mathrm{T}}=T_{\mathrm{TC}}$ and $%
T_{\mathrm{C}}\leq T_{\mathrm{TC}}$. In the T phase we set $P=0,$ $Q=Q_{0}$
in Eq. (\ref{chi-tilt}), with $Q_{0}$ found from the equilibrium condition (%
\ref{eq-Q}):

\begin{eqnarray}
0 &=&\frac{\partial G}{\partial Q}=\beta _{2}Q+\beta _{4}Q^{3}\rightarrow \nonumber \\
Q_{0}^{2} &=& -\beta _{2}/\beta _{4}=\beta _{0}/\beta _{4}\left( T_{\mathrm{T}%
}-T\right)
\end{eqnarray}%
and Eq. (\ref{chi-tilt}) becomes%
\[
\chi =\left[ \alpha _{2}+\gamma Q_{0}^{2}\right] ^{-1}=\frac{1}{\alpha
_{0}\left( T-T_{\mathrm{C}}\right) +\gamma \beta _{0}/\beta _{4}\left( T_{%
\mathrm{T}}-T\right) }
\]

or%
\begin{eqnarray}
\epsilon &=&1+\chi =1+\frac{1}{\left( T-T_{\mathrm{C}}\right) /C+B\left( T_{%
\mathrm{T}}-T\right) }  \label{chi-tilt0} \\
C &=&1/\alpha _{0}  \nonumber \\
B &=&\gamma \beta _{0}/\beta _{4}  \nonumber
\end{eqnarray}%
The FE transition occurs at the temperature $T_{\mathrm{FE}}$ for which $%
\chi $ diverges, and therefore the denominator is null:
\[
T_{\mathrm{FE}}=\frac{T_{\mathrm{C}}-B~C~T_{\mathrm{T}}}{1+B~C}
\]%
so that a cooperative (competitive) interaction with the tilt mode raises
(lowers) $T_{\mathrm{FE}}$ with respect to $T_{\mathrm{C}}$.

\subsubsection{Integration into the Lorentz--Lorenz formula}

The above expression of the permittivity, Eq. (\ref{chi-tilt0}), is obtained
neglecting the electronic and atomic contributions to the polarizabilities $%
\alpha _{el}$ and $\alpha _{a}$, namely setting $\epsilon _{\infty }=1$, as
usual in the context of FE with very large $\epsilon \simeq \chi $. In the
present case, however, this approximation is not well satisfied, and it is
better to insert the correction for polar-tilt coupling into the
Lorentz--Lorenz formula,\cite{Hip54,Raj03} which is the Clausius-Mosotti
formula with the inclusion of all types of polarizabilities
\[
\frac{\epsilon -1}{\epsilon +2}=\frac{N}{3\epsilon _{0}}\left( \alpha
_{el}+\alpha _{a}+\alpha _{\mu }\right) =\frac{\epsilon _{\infty }-1}{%
\epsilon _{\infty }+2}+\frac{N\alpha _{\mu }}{3\epsilon _{0}}\;
\]%
which can be rewritten as
\[
\frac{\epsilon -1}{\epsilon +2}-\frac{\epsilon _{\infty }-1}{\epsilon
_{\infty }+2}=\frac{N\mu ^{2}}{9\epsilon _{0}k_{\text{B}}T}=\frac{T_{0}}{T}
\]%
and, after some manipulation, as\cite{Fio71}%
\begin{eqnarray}
\epsilon &=&\epsilon _{\infty }+\frac{\left( \epsilon _{\infty }+2\right)
^{2}T_{0}}{3T-\left( \epsilon _{\infty }+2\right) T_{0}}=\epsilon _{\infty }+%
\frac{\left( \epsilon _{\infty }+2\right) T_{\mathrm{C}}}{T-T_{\mathrm{C}}}
\label{LL} \\
T_{\mathrm{C}} &=&\frac{\left( \epsilon _{\infty }+2\right) }{3}T_{0}=\frac{%
\left( \epsilon _{\infty }+2\right) N\mu ^{2}}{27\epsilon _{0}k_{\text{B}}}
\nonumber
\end{eqnarray}%
Comparing Eq. (\ref{LL}) and (\ref{chi-tilt0}) we set the Curie constant
\[
C=\left( \epsilon _{\infty }+2\right) T_{\mathrm{C}}
\]%
and insert the correction for tilt-polar coupling, $\left( T-T_{\mathrm{C}%
}\right) \rightarrow \left( T-T_{\mathrm{C}}\right) +B~C\left( T_{\mathrm{T}%
}-T\right) $, into Eq. (\ref{LL}), finding finally
\[
\epsilon =\epsilon _{\infty }+\frac{C}{T-T_{\mathrm{C}}+B~C\left( T_{\mathrm{%
T}}-T\right) }~.
\]


\end{document}